\documentclass[runningheads]{cl2emult}

\usepackage{makeidx}  % allows index generation
\usepackage{graphicx} % standard LaTeX graphics tool
\usepackage{subeqnar} % subnumbers individual equations
\usepackage{multicol} % used for the two-column index
\usepackage{cropmark} % cropmarks for pages without
\usepackage{eso}      % placeholder for figures
\makeindex            % used for the subject index
\begin{document}
\title*{Black hole states and radio jet formation\footnote{To be
published in `Black Holes in binaries and galactic nuclei'}}
\toctitle{Black hole states and radio jet formation}
% allows explicit linebreak for the table of content
%
%
\titlerunning{Black hole states and radio jet formation}

% allows abbreviation of title, if the full title is too long
% to fit in the running head
%
\author{R.P. Fender}
\authorrunning{R.P. Fender}
% if there are more than two authors,
% please abbreviate author list for running head
%
%
\institute{
Astronomical Institute `Anton Pannekoek' and Center for High Energy
Astrophysics, University of Amsterdam, Kruislaan 403, \\ 
1098 SJ Amsterdam, The Netherlands}

\maketitle              % typesets the title of the contribution

\begin{abstract}

An empirical relation between `canonical' X-ray states in black hole
X-ray binaries and radio emission is presented. In the Low/Hard state
a quasi-continuous radio-emitting jet is produced. In the the
High/Soft state the jet is not observed. In the Very High state and
at major state transitions, which may correspond physically to rapid
and/or significant changes in the inner accretion disc radius,
discrete ejection events, which manifest themselves as radio flares,
are observed. These relations appear to hold for both transient and
persistent systems. Furthermore it is argued that the extremely strong
coupling between hard X-rays and radio emission from black hole X-ray
binaries implies that the Comptonising corona is simply the base of
the jet.

\end{abstract}

\section{Introduction : `Canonical' black hole states}

Black hole candidate (BHC) X-ray binaries (Tanaka \& Lewin 1995) are
believed to consist of an accreting black hole with a mass in the
range $3$M$_{\odot} \leq M_{\rm BH} \leq 20$M$_{\odot}$ (see
e.g. Charles 1999 for a review of the dynamical evidence for black
holes in these systems). Four BHC systems in our Galaxy (Cyg X-1, GX
339-4, 1E1740.7-2942 and GRS 1758-258) and two in the LMC (LMC X-1 and
LMC X-3) are persistent X-ray sources, in the sense that their X-ray
emission is more or less constant over time (which means, in effect,
over several decades since their discoveries).  Several more systems,
the soft X-ray transients (SXTs), are known only to have undergone a
handful of X-ray outbursts during the history of X-ray astronomy, but
it is these systems which offer the most convincing dynamical evidence
for black holes. Charles (1999) lists several such systems, which include A
0620-00, GRO J0422+32, GS 2000+25 and GRS 1124-68. Generally speaking,
these X-ray outbursts of the SXTs are accompanied by radio outbursts
(Hjellming \& Han 1995; Kuulkers et al. 1999).

It was initially realised that black holes displayed at least two
spectral states, `high' (in this paper High/Soft) and `low'
(Low/Hard), based upon the strength of their soft ($\leq$ few keV)
X-ray emission (Tanaka \& Lewin 1995). An `Off' state was first
reported for GX 339-4 by Markert et al. (1973).  Additionally, the
`Very High' (Miyamoto et al. 1991) and `Intermediate' (M\'endez \& van
der Klis 1997) states have been identified. M\'endez,
Belloni \& van der Klis (1998) have shown that even unusual systems
like the `superluminal' transient GRO J1655-50 display these
`canonical' states.

The physical interpretation of the different states is based upon an
origin for the soft X-ray component in an accretion disc and the hard
X-ray component via Comptonisation of softer photons in a corona of
high-energy electrons. The most popular current models invoke a
truncated accretion disc in the Low/Hard X-ray state, inside of which
may be an advection-dominated flow (e.g. Esin et al. 1998), and the
presence of a strong Comptonising corona. In the High/Soft state the
disc component dominates and the corona is smaller and cooler. In the
Very High state both the disc and corona are strong; the Intermediate
state may be similar but at a lower overall luminosity.

\section{Radio emission and black hole state}

\subsection{State transitions}

How does radio emission relate to black hole state
transitions? For the transient sources the radio emission is
generally (although not always) associated with one or more discrete
ejections, the first (and generally foremost) of which is associated
with the rapid rise (corresponding to something like a
Off$\rightarrow$High/Soft state transition in the space of a day or
so) of X-ray emission at the time of the outburst. The evidence for
discrete ejections comes from both direct imaging and the radio light
curves (discrete peaks) and radio spectra (rapidly evolving to
optically thin and subsequently decaying) observed (e.g. Kuulkers et
al. 1999). As well as the transients, the persistent sources Cyg X-1
(Zhang et al. 1998) and maybe also GX 339-4 (Corbel et al. in prep)
show flaring at times of state transitions.

\subsection{Off and Low/Hard states}

How does radio emission relate to canonical state outside of state
transitions? We know from the X-ray transients that radio emission is
pretty weak (often undetectable)
during X-ray quiescence $\equiv$ Off state.
This is supported by weak radio detections of GX 339-4 when at very
low X-ray flux levels (Corbel et al. in prep; Fig 1).

Cyg X-1 and GX 339-4 are generally observed in the Low/Hard X-ray
state, and reveal very similar characteristics, namely weak (few mJy
at cm wavelengths) but steady radio emission with a flat/inverted
radio spectrum (i.e. spectral index $\alpha = \Delta \log S_{\nu} /
\Delta \log \nu \geq 0$)
which is correlated with both soft and hard X-ray
emission over approximately a factor of two in intensity (Brocksopp et
al. 1999 and references therein; Hannikainen et al. 1998; Corbel et
al. in prep). The radio emission from Cyg X-1 is additionally
modulated at the 5.6-day orbital period (Pooley, Fender \& Brocksopp
1999; Brocksopp et al. 1999), probably due to free-free absorption in
the dense stellar wind of the massive companion star.  Models for
flat/inverted radio emission generally invoke a partially
self-absorbed outflow of the kind originally envisaged for AGN cores
by Blandford \& K\"onigl (1979). This is supported by the large size
scale required (a sphere with radius greater than the binary separation of
Cyg X-1 is required to generate the observed cm-wavelength emission).
Final confirmation of the jet hypothesis has come from VLBA
observations of an extension on mas-scales from Cyg X-1 (Stirling et
al. 1997; de la Force et al. in prep). 

\begin{figure}[t]
\centering
\includegraphics[width=.8\textwidth, angle=270]{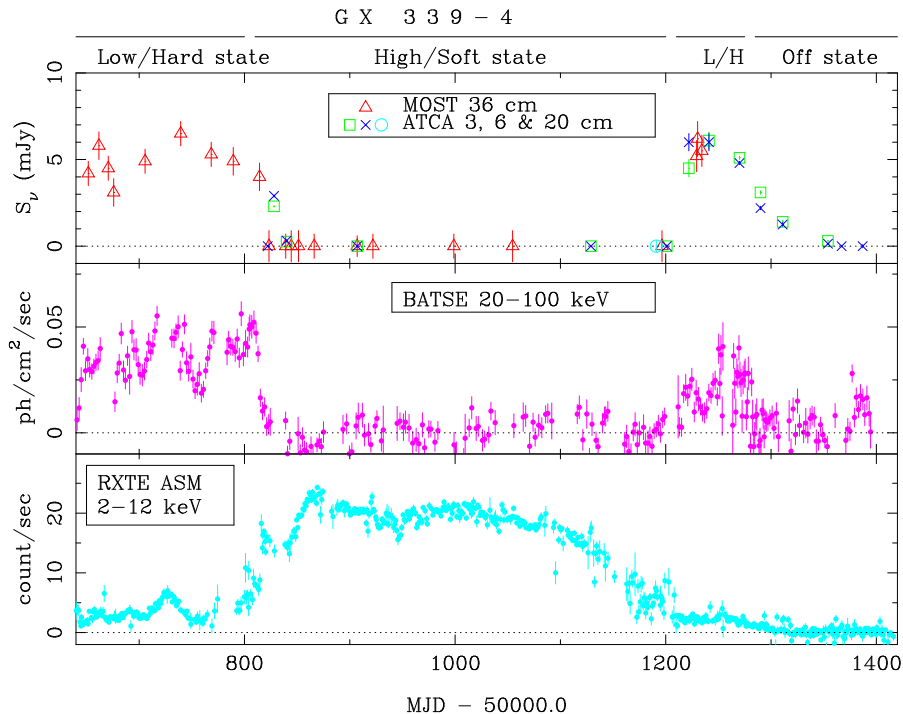}
\caption[]{Radio, soft- and hard-X-ray observations of the Low/Hard,
High/Soft and Off states in GX 339-4. From Fender et al. (1999) and
Corbel et al. (in prep).}
\label{eps1}
\end{figure}

As well as the persistent sources Cyg X-1 and GX 339-4, some X-ray
transients also spend extended periods in the Low/Hard state
(e.g. GRO J0422+32, GS 2023+338/V404 Cyg). These systems also
display a flat radio spectrum with comparable luminosity to Cyg X-1
and GX 339-4, when in the Low/Hard state (Fender, in prep).

\subsection{The High/Soft state}

It was already suspected that the radio flux density in Cyg X-1
dropped significantly when the system was in the High/Soft X-ray
state.  Dramatic confirmation that this was a phenomenon shared by
other BHCs came from long-term monitoring of GX 339-4 at radio, soft-
and hard-X-ray energies, in which it was found that the radio emission
from the system was reduced by a factor $\geq 25$ during a year-long
period in the High/Soft state (Fender et al. 1999). Figure 1
demonstrates this result, encompassing Low/Hard, High/Soft, Low/Hard
and finally Off states during the period of our monitoring program
(Corbel et al. in prep), and dramatically reveals the extremely strong
correlation between hard X-ray and radio emission. Is the outflow
physically suppressed during this state or cooled (via inverse Compton
losses) so rapidly by the more luminous disc that the electrons no
longer produce synchrotron emission ? The answer to this is unclear,
but at least one (naive) argument, the evidence for radio emission
during the (even more luminous) Very High State (see below), suggests
physical suppression of the jet.

Note that while many X-ray transients appear, at face value, to be
radio-bright whilst in the High/Soft state, more careful scrutiny
reveals that the radio emission is usually just the decaying tail
of emission from the flare event at the start of the outburst
(i.e. at the state transition), which is by then physically decoupled
from the system.

\subsection{The Very High (and Intermediate?) state(s)}

The Very High State is much rarer than the Low/Hard or High/Soft
states, and few clear examples exist of radio observations of this
state. GX 339-4, while having previously entered this state (Miyamoto
et al. 1991), has not done so since radio monitoring began.  GS
1124-683 also spent an extended period in the Very High State
(Miyamoto et al. 1993), during which time there {\em was} radio
coverage (for at least 30 days -- Kuulkers et al. 1999 and references
therein), revealing fairly bright and variable emission. However, the
radio coverage does not appear to have been good enough to distinguish
between rapid radio variability (as in GRS 1915+105, see below), or
simply two large, discrete ejections.

Belloni (1998) has suggested that GRS 1915+105, when exhibiting its highly
variable `dipping' behaviour, may be in the Very High State. This
pattern of behaviour is associated with bright and repeated flare
events with a flat spectrum from the radio -- mm -- infrared, which
are most likely associated with discrete ejection events $\equiv$
transient jet formation (Fender \& Pooley 1998 and references
therein). The intermediate state may be similar to the Very
High State, but at lower luminosity levels, but again there has been
no good radio coverage of this state.

\begin{figure}[t]
\centering
\includegraphics[width=.8\textwidth, angle=0]{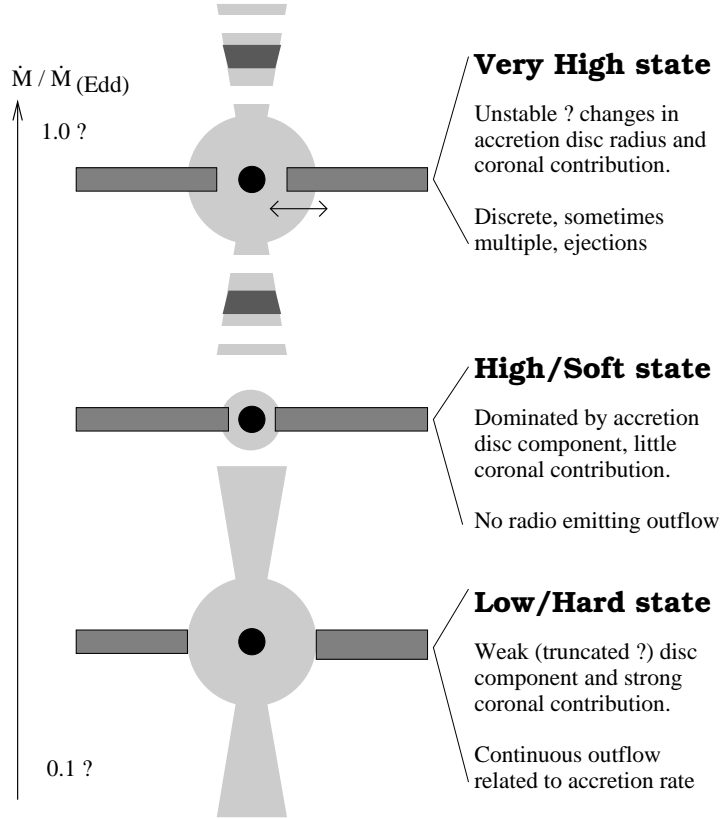}
\caption[]{Relation between models for BHC X-ray states and observed
radio emission}
\label{eps1}
\end{figure}

\section{Discussion}

\begin{table}
\caption{The relation of radio emission to black hole X-ray
state.}
\begin{tabular}{cc}
\hline
State & Radio properties \\
\hline
Very High & Bright ejections, spectral
evolution from absorbed $\rightarrow$ optically thin \\
High/Soft & Radio suppressed by factor $\geq 25$ \\
Intermediate & ? \\
Low/Hard & Low level, steady, flat spectrum extending to at least
sub-mm \\
Off       & Weak; similar to Low/Hard but reduced $\propto \dot{m}$ \\
\hline
\end{tabular}
\end{table}

We are now in a position to be able to characterise the relation of
radio emission from BHC X-ray binaries to their `state' as defined
observationally by their X-ray spectral and timing properties, except
for the Intermediate state, for which there is no clear case of
simultaneous radio observations. This relation is summarised in Fig 2
and Table 1. Several implications are quite clear

\begin{itemize}
\item{The Low/Hard state supports a quasi-continuous outflow whose
contribution to the overall energetics of the system is likely to be
significant. In this state there is a three-way correlation between
radio, soft- and hard-X-rays, which may reflect small changes in the
accretion rate. The Off state may simply be the Low/Hard state `turned
down' to lower accretion rates.}
\item{In the High/Soft state the radio emission drops below detectable
levels, probably corresponding to the physical disappearnce of the
jet. It is this state, in both radio and X-ray emission, which is most
different to the other states.}
\item{Rapid changes in the accretion disc radius (or whatever it is
that is physically changing which is currently modelled as an inner
disc radius!) correspond to discrete ejections, which appear as radio
(and sometimes mm and infrared) flares. This mechanism may be
operating in analogous ways in both general state transitions and in
the Very High State.}
\item{There is an extremely strong correlation between radio and hard
X-ray emission. The radio emission has been {\em directly observed} to
arise in outflows and to originate in synchrotron emission from a
population of high-energy electrons. The hard X-ray emission is {\em
inferred} to arise via Comptonisation by a similar population of
electrons (albeit the low-energy tail of the
distribution). Furthermore both components, the jet and corona, are
believed to originate at the centre of the accretion disc, in the
vicinity of the black hole. By far the simplest interpretation
therefore is that the Comptonising corona, at least in the case of BHC
systems, is simply the base of the jet.}
\end{itemize}

A simple observational relation between black hole state and radio
emission, and hence accretion and jet formation, has now been
established for the BHC systes. Furthermore all the evidence points to
the Comptonising corona in these systems being physically related to
the presence of a jet. The next stage will be to investigate these
relations quantitatively.

\section*{Acknowledgements}

RPF would like to thank Mariano M\'endez, Eric Ford, Jeroen Homan and
Michiel van der Klis for stimulating discussions.

%INDEX%%%%%%%%%%%%%%%%%%%%%%%%%%%%%%%%%%%%%%%%%%%%%%%%%%%%%%%%%%%%%%%
\clearpage
\addcontentsline{toc}{section}{Index}
\flushbottom
\printindex
%%%%%%%%%%%%%%%%%%%%%%%%%%%%%%%%%%%%%%%%%%%%%%%%%%%%%%%%%%%%%%%%%%%%%

\end{document}